\documentclass[12pt]{iopart}
\usepackage{graphicx}
\usepackage{iopams}
\begin{document}

\title[A new mechanism of electric dipole spin resonance]{A new mechanism of electric dipole spin resonance: hyperfine coupling in quantum dots}

\author{E A Laird$^1$, C Barthel$^1$, 
E I Rashba$^{1,2}$, C M Marcus$^1$, M~P~Hanson$^3$ and A C Gossard$^3$} 
\address{$^1$ Department of Physics, Harvard University, Cambridge, Massachusetts
02138, USA} 
\address{$^2$ Center for Nanoscale Systems, Harvard University, Cambridge, Massachusetts
02138, USA} 
\address{$^3$ Materials Department, University of California at Santa Barbara,
Santa Barbara, California 93106, USA} 
\ead{marcus@harvard.edu} 
\begin{abstract} 
A recently discovered mechanism of electric dipole spin resonance, mediated by the hyperfine interaction, is investigated experimentally and theoretically.  The effect is studied using a spin-selective transition in a GaAs double quantum dot.  The resonant frequency is sensitive to the instantaneous hyperfine effective field, revealing a nuclear polarization created by driving the resonance.  A device incorporating a micromagnet exhibits a magnetic field difference between dots, allowing electrons in either dot to be addressed selectively.  An unexplained additional signal at half the resonant frequency is presented.

\end{abstract} 
\submitto{\SST} 
\maketitle

\maketitle

\section{Introduction}

Electric dipole spin resonance (EDSR) is a method to electrically manipulate electron spins. In this technique, two fields are applied; a static magnetic field $\mathbf{B}$ and an oscillating electric field $\mathbf{\tilde{E}}(t)$ resonant with the electron precession (Larmor) frequency~\cite{RashbaEDSR, BellEDSR, McCombeEDSR, RashbaShekaBook}.  Spin resonance techniques are of interest for quantum computing schemes based on single electron spins, because they allow arbitrary one-qubit operations~\cite{LossDiVincenzo}.  Single-spin EDSR is a particularly desirable experimental tool because it allows spin manipulation without time-dependent magnetic fields, which are difficult to generate and localize at the nanoscale~\cite{JelezkoESR, RugarESR, XiaoESR,KoppensESR}.

Achieving EDSR requires a mechanism to couple $\mathbf{\tilde{E}}$ to the electron spin \mbox{\boldmath$\sigma$}.  This coupling can be achieved by the traditional spin-orbit interaction, which couples \mbox{\boldmath$\sigma$} to the electron momentum $\mathbf{k}$, or by an inhomogeneous Zeeman interaction, which couples \mbox{\boldmath$\sigma$} to the electron coordinate $\mathbf{r}$~\cite{PekarRashba, RashbaShekaBook,  KatoGTMR, TokuraSlanting, GolovachLoss}.  Single-spin EDSR has recently been achieved in quantum dots using both techniques~\cite{NowackEDSR, PioroLadriereEDSR}.

Recently, we presented an experimental  and theoretical study of a novel EDSR effect mediated by the spatial inhomogeneity of the hyperfine nuclear field~\cite{LairdEDSR}.  An electron moving under the influence of the electric field $\mathbf{\tilde{E}}(t)$ experiences this inhomogeneity as an oscillating hyperfine coupling which drives spin transitions.  In this paper, we illuminate the underlying physics and present new experimental data on a still unexplained phenomenon at half the resonant frequency.

This EDSR effect is observed via spin-blocked transitions in a few-electron GaAs double quantum dot~\cite{EngelESR}.  As expected for a hyperfine mechanism, but in contrast to the ${\bf k}-\mbox{\boldmath$\sigma$}$-coupling mediated EDSR, the resonance strength is independent of $\mathbf{B}$ at low field and shows, when averaged over nuclear configurations, no Rabi oscillations as a function of time.  We find that at large $\mathbf{B}$ driving the resonance creates a nuclear polarization, which we interpret as the backaction of EDSR on the nuclei~\cite{GueronOverhauser, DobersNMR, KoppensESR, BaughPolarization, RudnerEDSR}.  Finally, we demonstrate that spins can be individually addressed in each dot by creating a local field gradient.

\section{Device and measurement}

\begin{figure}
\centering \label{fig:fig1}
\includegraphics[width=8.5cm]{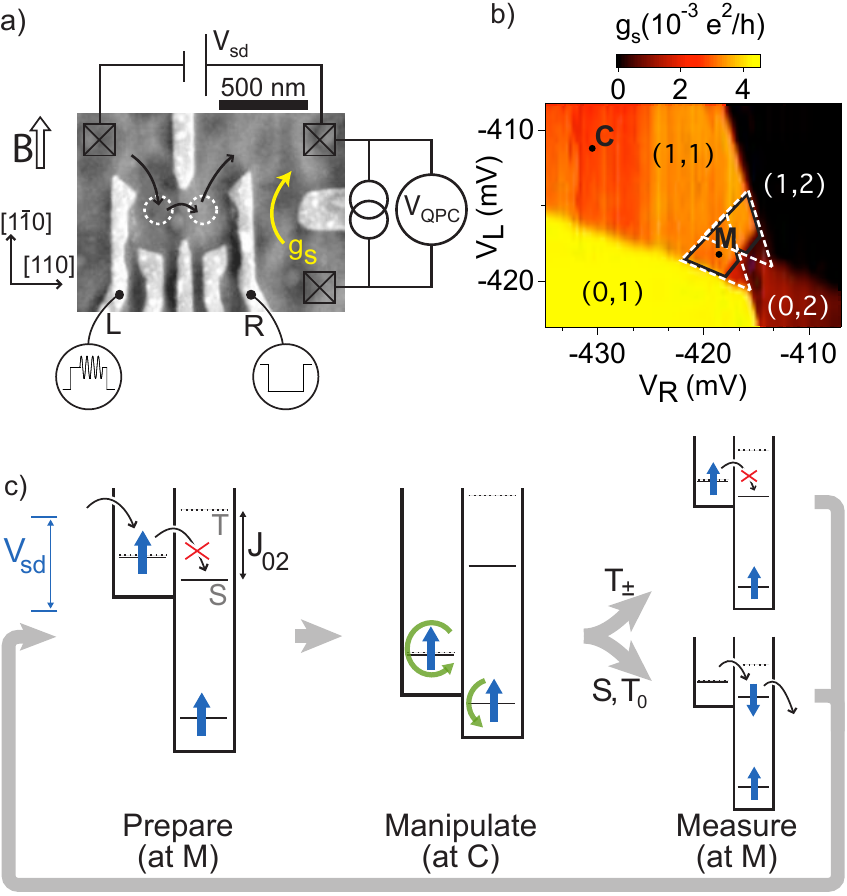}
\caption{(a) Micrograph of a device lithographically identical to the one measured, with schematic of the measurement circuit.  Quantum dot locations are shown by dashed circles, and a bias $V_\mathrm{sd}$ drives sequential tunneling in the direction marked by black arrows.  The conductance $g_{s}$ of the QPC on the right is sensitive to the dot occupation.  The direction of the magnetic field $\mathbf{B}$ and the crystal axes are indicated.  (b) QPC conductance $g_\mathrm{s}$ measured at $V_\mathrm{sd}\sim 600$~$\mu\mathrm{eV}$ near the (1,1)-(0,2) transition.  Equilibrium occupations for different gate voltages are shown, as are gate voltage configurations during the measurement/reinitialization (M)  and manipulation (C) pulses.   The two white dashed triangles outline regions where transport is not Coulomb blocked; the solid black line outlines where spin blockade is active.  A plane background has been subtracted.  (c) Energy levels of the double dot during the pulse cycle (See text).}
\end{figure}

The device for which most data is presented (Figure~1(a)) was fabricated on a GaAs/Al$_\mathrm{0.3}$Ga$_\mathrm{0.7}$As heterostructure with two-dimensional electron gas (2DEG) of density $2\times10^{15}$~m$^{-2}$ and mobility $20$ m$^2$/Vs located  110 nm below the surface. Voltages applied to Ti/Au top gates locally deplete the 2DEG, defining a few-electron double quantum dot. A nearby charge sensing quantum point contact (QPC) is sensitive to the electron occupation $(N_\mathrm{L},N_\mathrm{R})$ of the left $(N_\mathrm{L})$ and right $(N_\mathrm{R})$ dots~\cite{FieldSensing, Elzermansensing}.  The voltages $V_\mathrm{L}$ and $V_\mathrm{R}$ on gates L and R can be rapidly pulsed; in addition, L is coupled  to a microwave source.  The static magnetic field $\bf{B}$ was applied in the plane of the heterostructure, and measurements were performed in a dilution refrigerator at 150~mK electron temperature.

The characteristic feature of tunnel-coupled quantum dots is a discrete electron energy spectrum.  An overall shift to the spectrum, proportional to the electron occupation, is induced by $V_L$ and $V_R$, which therefore determine which occupation is energetically favoured.  Figure~1(b) shows the QPC conductance $g_\mathrm{s}$ as a function of $V_L$ and $V_R$; different conductances correspond to different ($N_\mathrm{L},N_\mathrm{R}$).  For most $V_L,V_R$ configurations, only one value of $(N_\mathrm{L},N_\mathrm{R})$ is energetically accessible; these correspond in Figure~1(b) to regions of uniform $g_s$.

A bias $V_\mathrm{sd}$ applied across the device drives electron transport via sequential tunneling subject to two constraints~\cite{HansonReview}.  The first constraint, Coulomb blockade, arises because for most gate configurations electrostatic repulsion prevents additional electrons from tunneling onto either dot.  This constraint inhibits transport except when $V_L,V_R$ are tuned so that three occupation configurations are near-degenerate.  The energy cost of an extra electron tunneling through the device is then small enough to be provided by the bias voltage.  Values of $V_L$ and $V_R$ satisfying this condition correspond to the two white dashed triangular regions marked in Figure 1(b), for which transport is permitted via the transition sequences $(0,2)\rightarrow(0,1)\rightarrow(1,1)\rightarrow(0,2)$ or $(0,2)\rightarrow(1,2)\rightarrow(1,1)\rightarrow(0,2)$.

A second constraint, spin blockade, is caused by the Pauli exclusion principle, which leads to an intra-dot exchange energy $J_{02}$ in the right dot~\cite{OnoSpinBlockade, JohnsonSpinBlockade}.  As shown in the first panel of Figure 1(c), the effect of this exchange is to make the $(1,1)\rightarrow(0,2)$ transition selective in the two-electron spin state, inhibited for triplet states but allowed for the singlet.  Although the hyperfine field difference between dots rapidly converts the $m_s=0$ component $T_0$ of the blocked triplet $T$ to an unblocked singlet $S$, decay of $m_s=\pm1$ components $T_\pm$ requires a spin flip and therefore proceeds much more slowly.  This spin flip becomes the rate-limiting step in transport, and so the time-averaged occupation is dominated by the (1,1) portion of the transport sequence \cite{JohnsonSpinBlockade}.  Gate configurations where spin blockade applies correspond to the black solid outlined region of Figure 1(b); inside this region, $g_\mathrm{s}$ has the value corresponding to (1,1).  Any process that induces spin flips will partially break spin blockade and lead to a decrease in $g_\mathrm{s}$.

Unless stated otherwise, EDSR is detected via changes in $g_\mathrm{s}$ while the following cycle of voltage pulses $V_\mathrm{L}$ and $V_\mathrm{R}$~\cite{KoppensESR} is applied to L and R (Figure~1(c)). The cycle begins inside the spin blockade region (M in Figure~1(b)), so that the two-electron state is initialized to $(1,1)T_\pm$ with high probability.  A $\sim$1~$\mu$s pulse to point C prevents electron tunneling regardless of spin state.  Towards the end of this pulse, a microwave burst of duration $\tau_\mathrm{EDSR}$ at frequency~$f$ is applied to gate L.  Finally the system is brought back to M for $\sim$3~$\mu$s for readout/reinitialization.  If and only if a spin (on either dot) was flipped during the pulse, the transition $(1,1)\rightarrow(0,2)$ occurs, leading to a change in average occupation and in $g_\mathrm{s}$.  If this transition occurs, subsequent electron transitions reinitialize the state to $(1,1)T_\pm$ by the end of this step, after which the pulse cycle is repeated.  This pulsed EDSR scheme has the advantage of separating spin manipulation from readout.

Changes in $g_\mathrm{s}$ are monitored via the voltage $V_\mathrm{QPC}$ across the QPC sensor biased at 5~nA.  For increased sensitivity, the microwaves are chopped at 227~Hz and the change in voltage $\delta V_\mathrm{QPC}$ is synchronously detected using a lock-in amplifier.  We interpret $\delta V_\mathrm{QPC}$ as proportional to the spin-flip probability during a microwave burst, averaged over the 100~ms lock-in time constant.

\section{EDSR spectroscopy}

\begin{figure}
\center \label{fig:fig2}
\includegraphics[width=8.5cm]{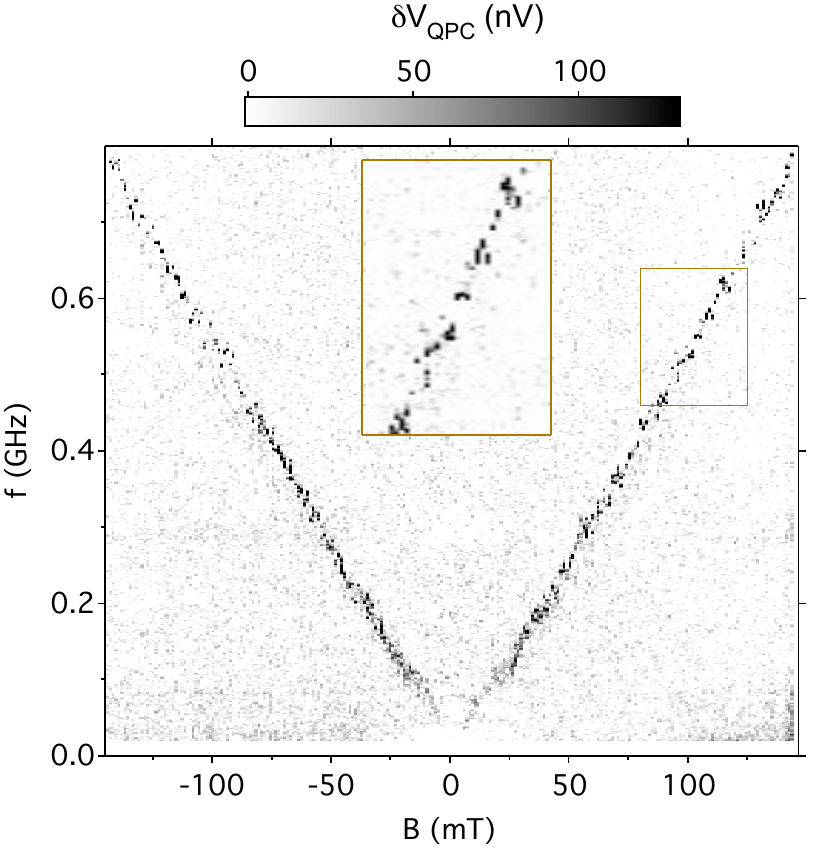}
\vspace{-0.3 cm} \caption{Signal of spin resonance $\delta V_\mathrm{QPC}$ as a function of magnetic field $B$ and microwave frequency $f$.  EDSR induces a breaking of spin blockade, which appears as a peak in the voltage across the charge sensor $\delta V_\mathrm{QPC}$ at the Larmor frequency.  Field- and frequency-independent backgrounds have been subtracted.  Inset: Jitter of resonant frequency due to random Overhauser shifts.}
\end{figure}

Resonant response is seen clearly as $B$ and $f$ are varied for constant $\tau_\mathrm{EDSR}=1~\mu$s (Figure~2.)  A peak in $\delta V_\mathrm{QPC}$, corresponding to a spin transition, is seen at a frequency proportional to $B$.  This is the key signature of spin resonance.  From the slope of the resonant line in Figure~2 a $g$-factor $|g|=0.39\pm0.01$is found, typical of similar GaAs devices~\cite{GGKondo, HansonZeeman}.  We attribute fluctuations of the resonance frequency (inset of Figure~2) to Overhauser shift caused by the time-varying hyperfine field acting on the electron spin.  Their range is $\sim\pm22$~MHz, corresponding to a field of $\sim$~4~mT, consistent with Overhauser fields in similar devices~\cite{KoppensNuclei, JohnsonT1, PettaT2}.

\begin{figure}
\center \label{fig:fig3}
\includegraphics[width=8.5cm]{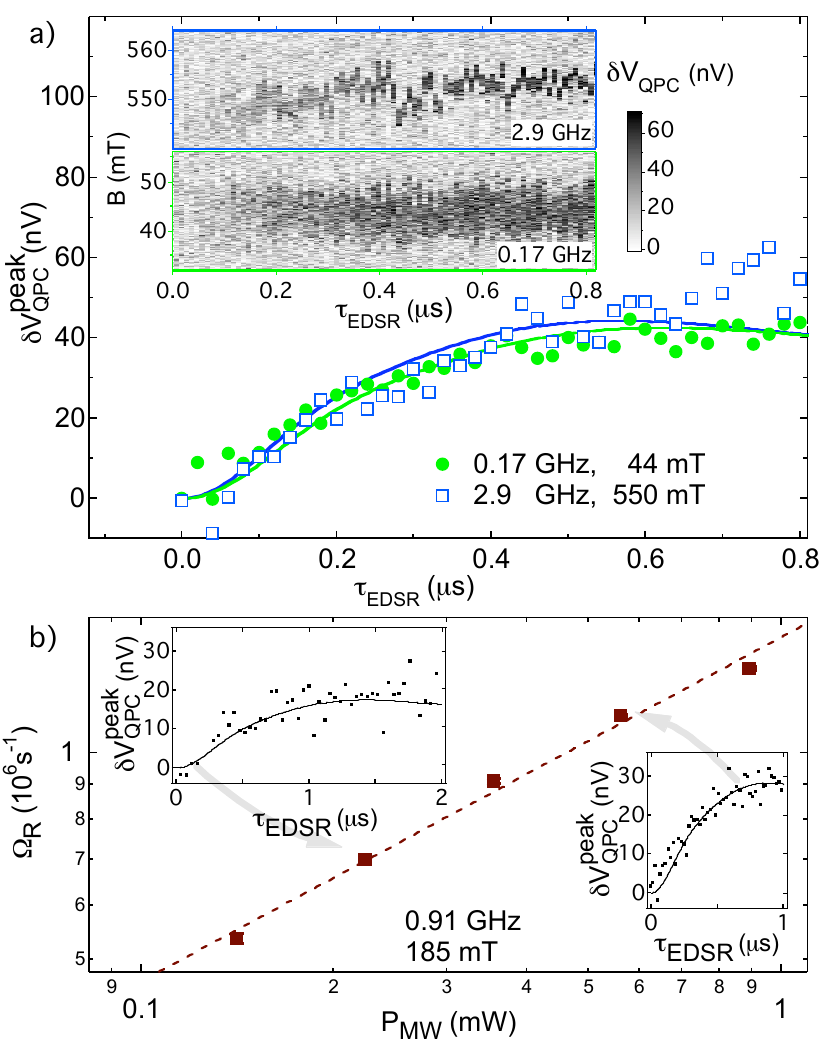}
\caption{(a) Measured EDSR peak strength $\delta V_\mathrm{QPC}^\mathrm{peak}$ (symbols) versus microwave pulse duration $\tau_\mathrm{EDSR}$ for two frequencies at equal power, along with theoretical fits (curves) obtained by numerically evaluating and scaling Equation~(4) (see text).  Inset: Raw data from which the points in the main figure are extracted.  Each vertical cut corresponds to one point in the main figure.  Jitter in the field position of the resonance reflects time-dependent Overhauser shifts.  (b) Spin-flip rate $\Omega_R$ as a function of applied microwave power $P_\mathrm{MW}$, along with a fit to the form $\Omega_R \propto \sqrt{P_\mathrm{MW}}$ (dashed line).  Insets: $\delta V_\mathrm{QPC}^\mathrm{peak}$ versus $\tau_\mathrm{EDSR}$ for two values of the microwave power, showing the fits from which points in the main figure are derived.}
\end{figure}

Information about the EDSR mechanism can be obtained by studying the peak height as a function of duration, strength, and frequency of the microwave burst (Figure~3).  To reduce the effects of the shifting Overhauser field, the microwave source is frequency modulated at 3~kHz in a sawtooth pattern with depth 36~MHz about a central frequency $\overline{f}$.  The resonance line as a function of  $\tau_\mathrm{EDSR}$ is shown in the inset of Figure 3(a).  For equal microwave power at two different frequencies $\overline{f}$, the peak heights  $\delta V^\mathrm{peak}_\mathrm{QPC}$ are plotted in Figure~3(a) (main panel).   The two data sets are similar in turn-on time and saturation value; this is the case for frequencies up to $\overline{f}=6$~GHz.  From similar data (insets of Figure~3(b)), using theory to be described, we extract the dependence of the spin-flip rate $\Omega_R$ on microwave power $P_\mathrm{MW}$ shown in the main panel of Figure~3(b).  Coherent Rabi-type oscillations in $\delta V^\mathrm{peak}_\mathrm{QPC}(\tau_\mathrm{EDSR})$ are not observed for any microwave power or magnetic field over the range measured.

The $B$-independence of the EDSR strength rules out spin-orbit mediated EDSR of the ${\bf k}-\mbox{\boldmath$\sigma$}$ type (either Dresselhaus or Rashba), for which the Rabi frequency is proportional to $B$~\cite{RashbaShekaBook, GolovachLoss, NowackEDSR}.  This is in contrast to the results of~\cite{NowackEDSR}, where the spin-orbit effect was found to dominate in a similar device to ours.  A possible explanation is the device orientation relative to $\mathbf{B}$ and the crystal axes.  In both our experiment and~\cite{NowackEDSR}, the gate geometry suggests a dominant $\bf{\tilde{E}}(t)$ oriented along one of the diagonal axes ([110] or [1$\overline{1}$0]), leading to an in-plane spin-orbit effective field $\bf{B}_\mathrm{eff}^\mathrm{SO}$ perpendicular to $\bf{\tilde{E}}(t)$. In our geometry (see Figure 1(a)), this orientation of $\bf{B}_\mathrm{eff}^\mathrm{SO}$ is parallel to $\bf{B}$, and therefore ineffective at driving spin transitions. In the geometry of~\cite{NowackEDSR}, $\bf{B}$ is perpendicular to $\bf{B}_\mathrm{eff}^\mathrm{SO}$, so that the ${\bf k}-\mbox{\boldmath$\sigma$}$ spin-orbit mechanism becomes more efficient .

\section{Theory}

A theoretical description of $\delta V^\mathrm{peak}_\mathrm{QPC}(\tau_\mathrm{EDSR})$ and its dependence on $B$ and $P_\mathrm{MW}$ can be obtained by modeling EDSR as arising from the coupling of an electron in a single dot to an oscillating electric field $\mathbf{\tilde{E}}(t)$ and the hyperfine field of an ensemble of nuclei~\footnote{There exists some physical similarity between the hyperfine mechanism of EDSR described in this paper and EDSR due to the coupling of electron spin to a random exchange field in semimagnetic semiconductors~\cite{KhazanEDSR}.}~\cite{LairdEDSR,RashbaEDSRTheory}.  Then the center of the dot oscillates as ${\bf R}(t)=-e{\bf \tilde{E}}(t)/m\omega_0^2$, where $m$ is the electron effective mass, and $\omega_0$ is its confinement frequency in a parabolic dot. As a result, the Hamiltonian of the hyperfine coupling of the electron spin $\mathbf{S}=\mbox{\boldmath$\sigma$}/2$ to nuclear spins $\mathbf{I}_j$ becomes time dependent, $H_{\rm hf}=A\Sigma_j\delta({\bf r}+{\bf R}(t)-{\bf r}_j)({\bf I}_j\cdot{\bf S})$. Here $A$ is the hyperfine coupling constant and the summation over $j$ runs over all nuclear spins. After expanding $H_{\rm hf}$ in ${\bf R}(t)$ (assumed small compared to the dot size) and averaging over the orbital ground-state wave function $\psi_0({\bf r})$ of the dot, the time dependent part of $H_{\rm hf}$ becomes $H_{\rm hf}(t)={\bf J}(t)\cdot\mbox{\boldmath$\sigma$}$, where ${\bf J}(t)$ is an operator in all ${\bf I}_j$. Choosing the $z$-axis in spin space along $\bf B$, the components of ${\bf J}(t)$ are $J_z={1\over2}A\sum_j\psi_0^2({\bf r}_j)I_j^z$ and

\begin{equation}
 J_\pm(t)={{eA}\over{m\omega_0^2}}\sum_j\psi_0({\bf r}_j){\bf \tilde{E}}(t)\cdot{\bf \nabla}\psi_0({\bf r}_j)I_j^\pm.
\label{eq1}
\end{equation}

  The time-dependent off-diagonal components $J_\pm(t)$ drive EDSR, while the quasi-static diagonal component $J_z$ describes detuning of EDSR from the Larmor frequency $\omega_L$ by an amount $\omega_z$ randomly distributed as $\rho(\omega_z)=\exp(-\omega_z^2/\Delta^2)/(\Delta\sqrt{\pi})$~\cite{MerkulovNuclei}. The dispersions $\Delta$ of the detuning and $\Omega_R$ of the Rabi frequency are the root-mean-square values of $J_z$ and $J_\pm$ respectively. Whereas $J_z$ is dominated by fluctuations of $\mathbf{I}_j$ symmetric about the dot centre, $J_\pm$ is dominated by fluctuations antisymmetric in the $\mathbf{\tilde{E}}$ direction because $\mathbf{\tilde{E}}\cdot\nabla\psi_0({\bf r})$ is odd with respect to the $\mathbf{\tilde{E}}$ projection of ${\bf r}$. Finally,
\begin{equation}
\Delta={{A}\over{2\hbar}}\sqrt{{{I(I+1)m\omega_0n_0}\over{2\pi\hbar d}}},\,
\Omega_R={{e\tilde{E}A}\over{\hbar^2\omega_0}}\sqrt{{{I(I+1)n_0}\over{8\pi d}}}\,,
\label{eq2}
\end{equation}
with $I=3/2$ for GaAs, $n_0$ the nuclear concentration, and $d$ the vertical confinement.  It is seen that $\Omega_R$ is independent of $B$; this is in contrast to EDSR mediated by the conventional ${\bf k}-\mbox{\boldmath$\sigma$}$ spin-orbit coupling, where Kramers' theorem requires that the Rabi frequency vanish linearly as $B\rightarrow0$~\cite{RashbaShekaBook, LevitovRashba, GolovachLoss}.

In an instantaneous nuclear spin configuration with detuning $\delta\omega=2\pi f-(\omega_L+\omega_z)$ and Rabi frequency $\Omega$, the spin-flip probability from an initial $\uparrow$ spin state is~\cite{RabiOscillations}:
\begin{equation}
p_\downarrow(\tau_\mathrm{EDSR})={{\Omega^2}\over{\left(\delta\omega/2\right)^2+\Omega^2}}
\sin^2{\left[\sqrt{\left(\delta\omega/2\right)^2+\Omega^2}~\tau_\mathrm{EDSR}\right]}\,.
\label{eq3}
\end{equation}
(We neglect the electron spin relaxation and nuclear-spin dynamics, which remain slow compared with the Rabi frequency even in the EDSR regime~\cite{PettaT2, RashbaEDSRTheory}.) To compare with the time-averaged data of Figure~3, we average Equation~(\ref{eq3}) over $\omega_z$ with weight $\rho(\omega_z)$ and over $\Omega$ with weight $\rho(\Omega)=2\Omega\exp(-\Omega^2/\Omega_R^2)/\Omega_R^2$. This latter distribution arises because the $J_\pm$ acquire Gaussian-distributed contributions from both $I^x_j$ and $I^y_j$ components of the nuclear spins, hence it is two-dimensional.  Averaging over $\omega_z$ and $\Omega$ results in a mean-field theory of the hyperfine-mediated EDSR. The resulting spin-flip probability
\begin{equation}
\overline {p}_\downarrow (\tau_\mathrm{EDSR};\Delta, \Omega_R) = 
\int_{-\infty}^{+\infty} d \omega_z \,  \rho(\omega_z) \int_{0}^{+\infty} d \Omega \,  \rho(\Omega) p_\downarrow(\tau_\mathrm{EDSR})
\end{equation}
shows only a remnant of Rabi oscillations as a weak overshoot at $\tau_\mathrm{EDSR}\sim\Omega_R^{-1}$.  The absence of Rabi oscillations is characteristic of hyperfine-driven EDSR when the measurement integration time exceeds the nuclear evolution time~\cite{ReillyNuclearEvolution}, and arises because~$J_\pm$ average to zero.

\subsection{Comparison with data}
To compare theory and experiment, the probability $\overline {p}_\downarrow(\tau_\mathrm{EDSR};\Delta, \Omega_R)$ is scaled by a QPC sensitivity $V^0_\mathrm{QPC}$ to convert to a voltage $\delta V_\mathrm{QPC}^\mathrm{peak}$. After scaling, numerical evaluation of Equation`(4) gives the theoretical curves shown in Figure~3(a). The parameters that determine these curves are as follows: The Larmor frequency spread, $\Delta=2\pi \times 28$~MHz, is taken as the quadrature sum of the jitter amplitude seen in Figure~2 and half the frequency modulation depth, whereas $\Omega_R$ and $V^0_\mathrm{QPC}$ are numerical fit parameters. The 44~mT data (green curve in Figure~3(a)) give $\Omega_R = 1.7\times10^6$ $\mathrm{s}^{-1}$ and $V^0_\mathrm{QPC}=2.4$~$\mu$V.  Holding $V^0_\mathrm{QPC}$ to this value, the 550~mT data give $\Omega_R=1.8\times10^6$~$\mathrm{s}^{-1}$ (blue curve in Figure~3(a)) and the 185~mT data give the dependence of $\Omega_R$  on microwave power $P_\mathrm{MW}$ shown in Figure~3(b).  The Rabi frequency $\Omega_R$ increases as $\sqrt{P_\mathrm{MW}}$ (Figure~3(b)) and is independent of $B$, both consistent with Equation~(1).  The $B$-independence of $\Omega_R$ --- also evident in the EDSR intensity in Figure~2---and the absence of Rabi oscillations support our interpretation of hyperfine-mediated EDSR in the parameter range investigated.

Estimating $\hbar\omega_0 \sim 1$ meV~\cite{HansonZeeman}, $\tilde{E}\sim 6\times10^3$ Vm$^{-1}$ at maximum applied power~\footnote{The maximum power is limited by non-resonant lifting of spin blockade, which we take to indicate a microwave amplitude exceeding the gate voltage from C to the nearest charge transition.  The data in Figure~3(a) and the last data point in Figure~3(b) use power $2\pm1$~dB below this threshold, corresponding to 3.2~mV.  Dropped uniformly across the whole device this voltage gives a field $\tilde{E}\sim3\times10^3$~Vm$^{-1}$.}, $d\sim5$ nm, and using values from the literature $n_0=4\times 10^{28}$~m$^{-3}$ and $An_0$=90~$\mu$eV \cite{PagetHyperfine} we calculate $\Omega_R\sim11\times10^6$ $\mathrm{s}^{-1}$, an order of magnitude larger than measured.  The discrepancy may reflect uncertainty in our estimate of $\tilde{E}$.

Above, we generalized a mean-field description of the hyperfine interaction~\cite{KhaetskiiNuclei, MerkulovNuclei} to the resonance regime. Justification for this procedure was provided recently in~\cite{RashbaEDSRTheory}. A distinctive feature of the mean-field theory is a weak overshoot, about 10 - 15\%, that is expected in the data of Fig. 3(a) before $\delta V _\mathrm{QPC}^\mathrm{peak}(\tau_\mathrm{EDSR})$ reaches its asymptotic value at $\tau_{\rm EDSR}\rightarrow\infty$. No overshoot is observed in the 550~mT data (blue symbols in Figure~3(a)), which was taken in a parameter range where an instability of the nuclear polarization begins to develop; see Section~5. For the 44~mT data (green symbols in Figure~3(a)), a considerable spread of experimental points does not allow a specific conclusion regarding the presence or absence of an overshoot. The theory of~\cite{RashbaEDSRTheory} suggests that the existence of the overshoot is a quite general property of the mean-field theory.  However, after passing the maximum, the signal decays to its saturation value vary fast, with Gaussian exponent $e^{-\Omega_\mathrm{R}^2 \tau_\mathrm{EDSR}^2}$.  By contrast, the first correction to the mean-field theory decays slowly, as $1/(N \Omega_\mathrm{R}^2 \tau_\mathrm{EDSR}^2)$, where $N$ is the number of nuclei in the dot.  As a result, the two terms become comparable at $\tau_\mathrm{EDSR} \sim \sqrt{\ln N}/ \Omega_\mathrm{R}$, which should make the maximum less pronounced.  Because for $N \sim 10^5$ the factor $\sqrt{\ln N} \sim 3$, the corrections to the mean-field theory manifest themselves surprisingly early, at times only about $\tau_{\rm EDSR}\approx3/\Omega_R$, making the overshoot difficult to observe.

\section{Nuclear polarization}

 \begin{figure}
\center \label{fig:fig4}
\includegraphics[width=8.5cm]{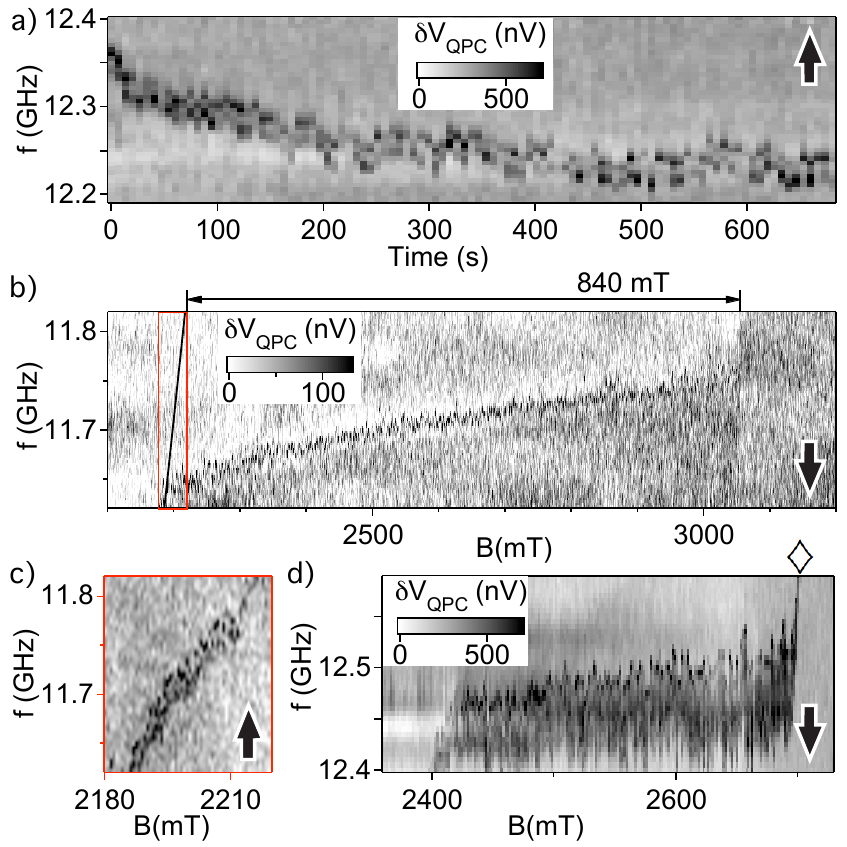}
\vspace{-0.3 cm} \caption{(a)  Shift of the resonance frequency with time at constant $B=2390$~mT, showing build-up of nuclear polarization over $\sim 200$~s. (b) A scheme to allow larger polarizations: the microwave frequency is repeatedly scanned over the resonance while $B$ is swept upwards.  Nuclear polarization partly counteracts $B$, moving the resonance away from its equilibrium position (black diagonal line) by up to 840 mT.  (c) Similar data taken at lower microwave power and opposite frequency sweep direction, showing approximately the equilibrium resonance position. (Grey scale as in (b)). (d) Similar data as in (b), with faster sweep rate, showing more clearly the displacement and subsequent return to equilibrium of the resonance.  $\diamondsuit$ marks the escape of the resonance from the swept frequency window.  In all plots, arrows denote frequency sweep direction.}
\end{figure}

Consistent with a hyperfine mechanism, this EDSR effect can create a non-equilibrium  nuclear polarization~\cite{BaughPolarization}.  If $f$ is scanned repeatedly over the resonance at high power, a shift of the resonance develops (Figure~4(a)), corresponding to a nuclear spin alignment parallel to $\mathbf{B}$.  The effect is stronger at higher $B$, and saturates over a timescale $\sim 200$~s.  In Figure~4(b), we show how to build up a substantial polarization:  While slowly increasing $B$, we scan $f$ repeatedly downwards, i.\,e., in the direction which tracks the moving resonance.  The resonance frequency remains approximately fixed, showing that the developing polarization compensates the increase in $B$. From the maximum line displacement from equilibrium, an effective hyperfine field of 840~mT can be read off, corresponding to a nuclear polarization of $\sim 16\%$.  Figure 4(c) shows similar data for lower power and opposite frequency sweep direction, indicating the approximate equilibrium line position.  Figure~4(d), similar to Figure~4(b) but with a faster sweep rate, makes the displacement and eventual escape of the resonance clearer although the maximum polarization is less.

The resonance shift is observed to be towards lower frequency, corresponding to a nuclear polarization parallel to $\mathbf{B}$.  This can be understood if the pulse cycle preferentially prepares the electron ground state $T_+$ over $T_-$, either because it is more efficiently loaded or because of electron spin relaxation.  EDSR then transfers this electron polarization to the nuclei~\cite{RudnerEDSR}.  We emphasize that the line shift is opposite to what is given by the usual Overhauser mechanism for inducing nuclear polarization via electron resonance~\cite{OverhauserPolarization, GueronOverhauser}.

\section{Addressing individual spins}

\begin{figure}
\center \label{fig:fig5}
\includegraphics[width=8.5cm]{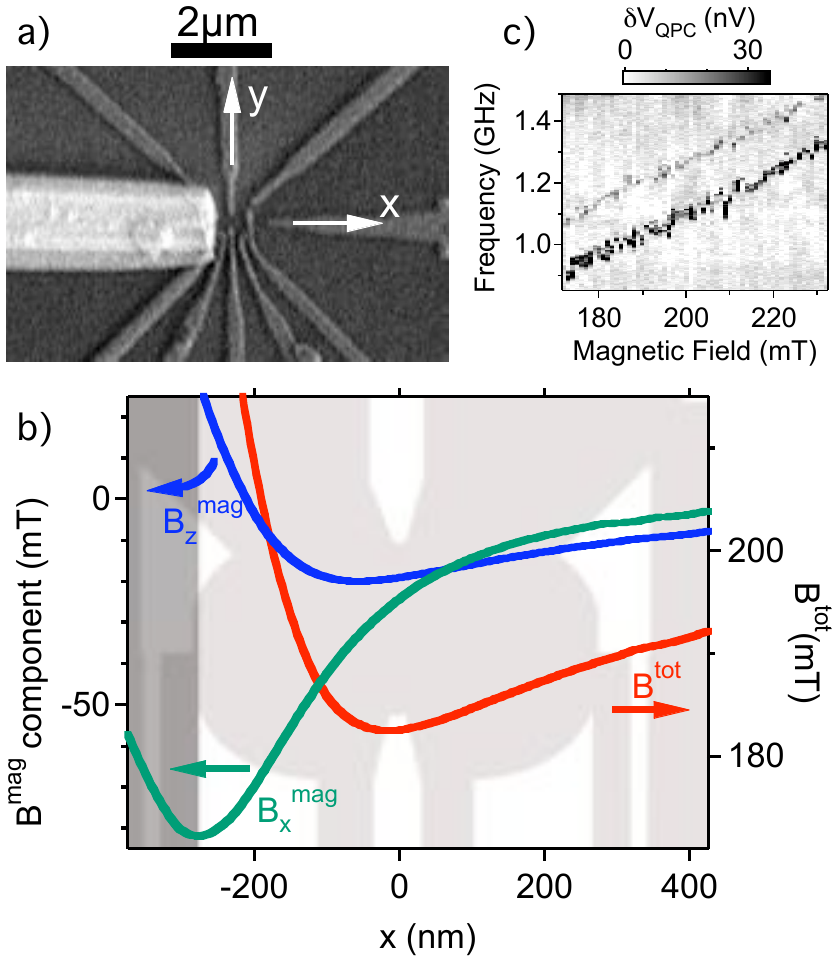}
\vspace{-0.3 cm} \caption{(a) A device similar to that of Figure~1, incorporating a micromagnet. (b) Total field magnitude $B^\mathrm{tot}$ (right axis) and the $x$ and $z$ components of the micromagnet contribution $\mathbf{B}^\mathrm{mag}$ (left axis), simulated at $y=0$ for external field $B=200$~mT along $\hat{\bf{z}}$ (out of the plane). $B^\mathrm{mag}_y$ vanishes by symmetry. The gate layout is shown in the background. (c) The associated split EDSR line. The lower resonance is stronger, as expected if the left electron is confined close to the minimum of $B_\mathrm{tot}$}.
\end{figure}

In quantum information applications, it is desirable to address individual spins selectively~\cite{LossDiVincenzo}.  A scheme to allow this is presented in Figure~5.  In an otherwise similar device~(Figure 5(a)), we incorporated a 100~nm thick micron-scale permalloy (84\% Ni, 16\% Fe) magnet over 35~nm of atomic-layer-deposited alumina~\cite{TokuraSlanting, PioroLadriere}.  This device was measured with external field $\mathbf{B}$ normal to the heterostructure plane.  A finite-element simulation of the field $\mathbf{B}^\mathrm{mag}$ due to the micromagnet, assuming complete permalloy magnetization along $\mathbf{B}$, yields the field profiles shown in Figure~5(b). The difference in total field $B^\mathrm{tot}=|\mathbf{B}+\mathbf{B}^\mathrm{mag}|$ between dots is $\sim 7$ mT.  As expected, the EDSR line measured in this device is frequently split (Figure~5(c)). The splitting, $10-20$~mT depending on precise gate voltage and pulse parameters, is not observed without the magnet and presumably reflects the field difference between dots.  Since this splitting is considerably larger than the Overhauser field fluctuations, spins in left and right dots can be separately addressed by matching $f$ to the local resonance condition~\cite{PioroLadriereEDSR}.

\section{Open issues and discussion}

\begin{figure}
\center \label{fig:fig6}
\includegraphics[width=8.5cm]{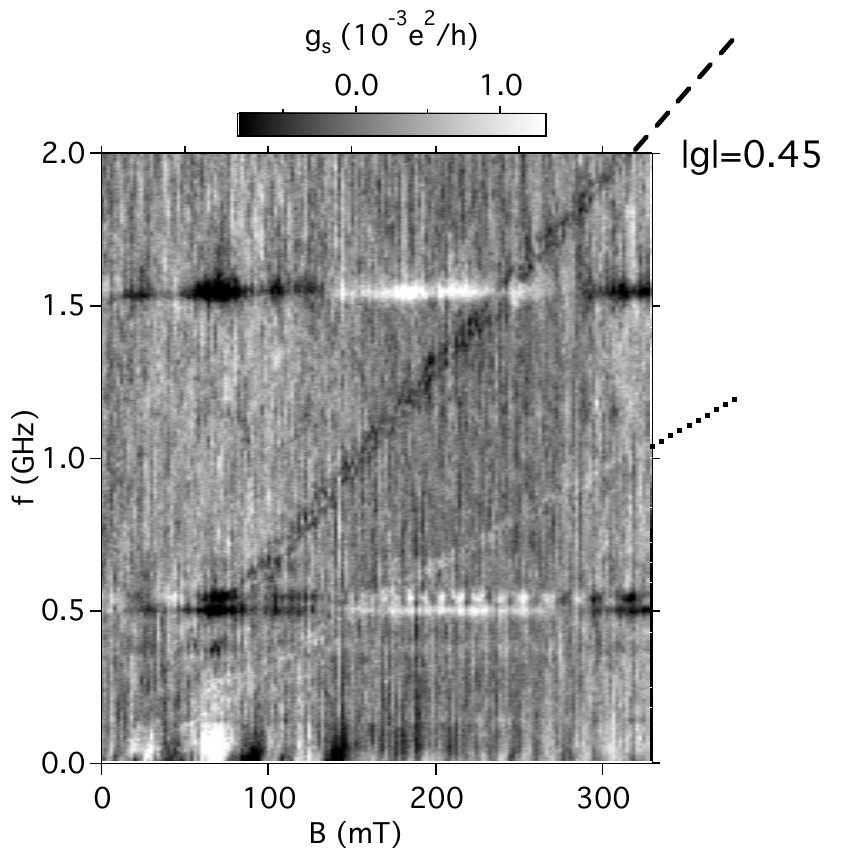}
\caption{Spin resonance signal (measured in conductance) in the device of Figure 5(a).  The EDSR signal shows up as a decrease in conductance as expected at frequency corresponding to $|g| = 0.45$ (marked with dashed line.)  An additional signal of opposite sign appears at exactly half this frequency (dotted line).  As in Figure 2, field- and frequency-independent backgrounds have been subtracted.  The horizontal features at 0.5 and 1.5~GHz result from resonances of the microwave circuit.}
\end{figure}

Finally, we discuss unexplained behavior observed only in the device of Figure~5(a).  For the data described in this section, a simplified measurement scheme is used:  Rather than applying gate pulses, the device is configured in the spin blockade region (point M in Figure 1(a)) throughout.  Microwaves are applied continuously, and spin resonance is detected by directly measuring the QPC conductance $g_\mathrm{s}$.

As well as the EDSR signal at full frequency $f=g\mu_B B/h$, an unexpected half-frequency signal is sometimes seen (Figure 6.)  Furthermore, depending on the exact gate configuration, both full-frequency and half-frequency signals can have either sign; the change in $g_\mathrm{s}$ at full frequency is usually negative as expected, but sometimes positive close to degeneracy of (1,1) and (0,2) charge configurations, where spin blockade is weakest~\cite{KoppensNuclei}; by contrast, the change in $g_\mathrm{s}$ at half frequency is usually positive but sometimes negative far from degeneracy.  For most gate configurations, full-frequency and half-frequency signals have opposite sign, as seen in Figure~6.

A half-frequency response is as far as we know unprecedented in spin resonance, and suggests second harmonic generation (SHG) from the microwave field.  SHG is generally a non-linear phenomenon; it occurs for example in optical materials with non-linear polarizability~\cite{FrankenSHG} and in non-linear electronic components such as diodes.  In our system, a hyperfine field at a harmonic of the microwave frequency arises if the confinement potential is non-parabolic.

However, SHG alone does not explain the sign of the conductance change seen at half-frequency in Figure~6.  A positive signal could in principle be caused by an admixture of the (0,1) charge state; but it is observed even for the gate configurations where (0,1) is energetically inaccessible (in the top right of the spin blockade region of Figure 1(b)).  Also, there is no reason why (0,1) should be admixed for one resonance but not the other.  These anomalous behaviours are therefore left unexplained.

\ack{We acknowledge useful discussions with Al.\ L. Efros, H.-A.~Engel, F. H. L.~Koppens, J. R. Petta, D. J. Reilly, M. S. Rudner, J. M. Taylor, and L. M. K. Vanderypen. We acknowledge support from the DTO and from DARPA. E.~I.~R. was supported in part by a Rutherford Professorship at Loughborough University, U.K.}

\end{document}